\documentclass[aps,twocolumn,prd,showpacs,nofootinbib]{revtex4}
\usepackage{amsmath}
\usepackage{graphicx}
\usepackage{dcolumn}
\usepackage{bm}
\usepackage{amssymb}
\usepackage{latexsym}

\def\be{\begin{equation}}
\def\ee{\end{equation}}
\def\ba{\begin{eqnarray}}
\def\ea{\end{eqnarray}}

\begin{document}

\title{ Inflation in a Web}

\author{Sheng Li$^{a}$, Yang Liu$^{a,b}$, Yun-Song Piao$^a$}
\affiliation{${}^a$ College of Physical Sciences, Graduate School
of Chinese Academy of Sciences, Beijing 100049, China}
\affiliation{${}^b$ Department of Physics, Shandong University,
Jinan, 250100, China}

\begin{abstract}

In a given path with multiple branches,
in principle, it can be expected that there are some fork points,
where one branch is bifurcated into different branches, or various
branches converge into one or several branches. In this paper, it
is showed that if there is
a web formed by such branches
in a given field
space, in which each branch can be responsible for a period of
slow roll inflation,
a multiverse separated by domain wall network will come into
being, some of which might corresponds to our observable universe.
We discuss this scenario and show possible observations of a given
observer at late time.

\end{abstract}


\maketitle

\section{Introduction}

Recently, the string landscape \cite{KKLT},\cite{Susskind} with
large number of vacua has received increasing attentions. In the
low energy limit, the landscape can be visualised as a complicated
and rugged potential in a given field space with multiple
dimensions.
The large number of extrema in the landscape means there is an
exponentially large number of paths for going down hill, some of
which may be suitable candidates for inflation
\cite{Guth},\cite{Linde82},\cite{AS},\cite{Starobinsky}.

In a randomly chosen downhill path, which generally has
zigzags and turns as a result of the complexity of the landscape,
there preferable be multiple stages of inflation, each of which is
dominated by the effective potential respectively.
The notion that inflation with multiple stages e.g.
\cite{PS},\cite{Feng03},\cite{DK},\cite{E},\cite{BEM3},\cite{LPS}
is appealing, since it can lead to a significant scale dependence
of spectral index of curvature perturbation.
However, this time we will not merely focus on a single
downhill path as commonly investigated but on the overall configuration
of the interveined paths.
Specifically speaking, there should be certainly lots of
fork points in a given path of field space, where one branch is
bifurcated into different branches, or various branches converge
into one or several branches\footnote{where the word ``stage'' is
replaced with ``branch'', which can be more visual for our purpose.}.
In this sense, this corresponds that there is a ``web'' formed by
connected branches in which each realistic path is consisted of a
series of conjoint branches.
Considering what if the inflation is driven by such a
web, which is our purpose of the paper, might be interesting since
conventionally the background of inflating universe is homogenous
all along for slow rolling inflaton, however, here since the fork
points inevitably alter the experience of different regions of
inflating universe, such homogeneousity can be hardly preserved in
whole space.
Therefore, we need to check the global configuration of inflating
universe under such a web.

When the effective inflaton field arrives at a given fork point
(here the effective inflaton means the effective field moving along
certain branch of a given path in field space, which can be one or
the synthesis of several fields) the inflating universe will turn
into many independent regions separated by domain walls \footnote{
Here, the fork point is slightly similar to the `waterfall' point
in hybrid inflation \cite{Linde94}, also
\cite{AF},\cite{LL},\cite{CLL},\cite{S}. The appearance of domain
walls is a main problem of original hybrid inflation, which needs
to be avoided, see \cite{LP},\cite{LS},\cite{GLW}. Here, however,
we identify the regions inside domain walls as different
universes, which experience the slow roll inflation till the
effective inflaton meets the fork point again or reheats.}.
The probability that the effective inflaton enters into certain
branch is determined by the characters of potential around the
corresponding fork point and the history.
After this splitting, the evolutions of different regions
will be generally different as the effective inflaton in each new
branch has different potentials.
In this sense, each of different regions actually corresponds to a
new universe, which can be stretched exponentially during its following
inflation. When the effective inflaton meets the fork again, the
separation will inevitably occur again. Thus it can be seen that
the existence of such a web of branches will lead to a multiverse
separated by domain wall network, especially,
discriminate from multiverse scenario, this multiverse is complemented
by the classical rolling of inflaton rather than the random walk by
its large quantum fluctuation and therefore takes place at a relatively
lower energy scale.
We will discuss this scenario,
and show possible observations of a given observer at late time,
if our universe is just one of this multiverse system.

\section{Description of Scenario}

We begin with a potential plotted in Fig.1, in which a simple web
of branches is given, and each of these branches is assumed to be
able to give a period of slow roll inflation. In general, when the
scale of potential is sufficiently high the fluctuation of field
can be expected to overwhelm its classical evolution. In this
case, the effective inflaton field will be in stochastic walking
with the step length ${H_{inf}\over 2\pi}$, where $H_{inf}$ is the
inflation scale. In unit of time ${1\over H_{inf}}$, different
regions with length scale ${1\over H_{inf}}$ will has different
energy density and thus denote different universes. This scenario
has been called as eternal inflation \cite{vilenkin},\cite{linde},
and also \cite{S1983},\cite{GW}, which leads to an inflationary
multiverse. In some universes of this inflating multiverse, the
field will fluctuate up and the selfproduction of universes
continues. While in
some other universes the field will
fluctuate downhill and into a regime where the classical evolution
dominates and the slow roll inflation will appear.
This eternal inflation in certain region could actually guarantee
that the web which is downriver of it will eventually get throughout
covered. In general, it
is thought for the latter that a reheating will occur and then the
corresponding universe will enter into an evolution of standard
cosmology. Here, however, we will show that in a downhill path,
due to the possible existence of some fork points, the case will
not so simple.

\begin{figure}[t]
\begin{center}
\includegraphics[width=8cm]{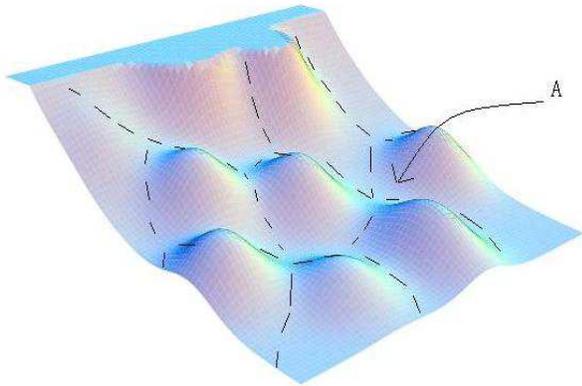}
\caption{The sketch of branches web. The dashed lines correspond
to the branches, along which the effective inflaton rolls down.
There are some fork points, where one branch is bifurcated into
different branches, or various branches converge into one or
several branches. }
\end{center}
\end{figure}

When the effective inflaton rolls down along its branch \footnote{
Hereafter, for our purpose we will be constrained to the
case that the fluctuation of field is not larger than its
classical evolution.}, it is bounded well in its orbit. While the
effective inflaton arrives at the fork point, e.g.`A', its orbit
will becomes an unstable ridge. 
The deviation of the extremum, which is
induced by the fluctuation of field
\cite{AAS},\cite{GLM},\cite{Linde92},\cite{Linde90}, along other
branches will bring the field off the ridge and 
make it roll
down to the corresponding branch. In the momentum space, the
fluctuation of field $\varphi$ is $\delta \varphi_k\simeq
{H_{\inf}\over \sqrt{2k^3}}$,
which means in time interval ${1\over
H_{inf}}$, the fluctuation in coordinate space is given by
${H_{inf}\over 2\pi}$, which occurs in different regions with
length scale ${1\over H_{inf}}$. The randomicity of fluctuation
means that each different ${1\over H_{inf}}$ region will be
expected to enter into different branch. In principle, the number
of new bifurcated branches at fork point should be large, since
the dimension of field space is quite large. In general, different
branches should have different potentials unless there is a large
fine tunning,
and the subsequent evolutions of different regions
are generally different. In this sense, each of these independent
regions actually corresponds to a new universe. These new
universes are generally separated by domain walls
\footnote{However, if the number of new bifurcated branches at
corresponding fork point is quite few, it is possible that there
are lots of adjacent $ {1\over H_{inf}}$ regions which will enter
into same branch. In this case, there will be not the domain walls
between them. }, which attributes to the complex structure of
effective potential.

The number of different regions, or universes, which are separated
by domain walls after `A', is given by the efolding number of
inflation along the branch before `A'. The universe with initial
length ${1\over H_{inf}}$ will become $R\simeq {1\over
H_{inf}}e^{\cal N}$ after ${\cal N}$ efoldings. Thus the number of
new universes after `A' is \be N_{\rm A}= \left({R\over ({1\over
H_{inf}})}\right)^3\simeq e^{3{\cal N}}, \ee which shows that the
larger the efolding number is, the more the number of new
universes after `A' would be. The inflation in each new universe will be
driven by the effective potential of corresponding new branch.
When the effective inflaton along this branch meets the fork point
again, a similar splitting process will appear.
In principle, this separation can occur for all new universes, each
of which undergoes an expansion driven by the inflaton along itself
branch. 
In this way, with the lapse of time, a multiverse actually comes into
being \footnote{In practice, there are certainly some fork points,
at which some branches converges into one branch. However, in this
case, generally it can hardly expected that the regions dominated
by these branches can be incorporated into one single universe,
since the experiences of these regions are certainly different
before their effective inflatons arrive at the fork point, unless
there is a large fine tunning for the effective potentials of
different branches. However, if there are only slight difference
between the effective potentials of corresponding branches, the
regions from different branches may be incorporated into one
observable universe, which can have distinct features in its CMB
power spectrum \cite{LW}. }.

In principle, it can be imagined that all paths could end at
certain stable or metastable minimum, in which reheating might
occur in some universes, lie with the potential in corresponding
branch, and then these universes will begin the evolution of
standard cosmology. The number of universes eventually entering
into certain stable or metastable minimum can be estimated as \be
N=\sum_{\rm all\,\, paths} \left(\prod_j f_{ij} e^{3{\cal
N}_{ij}}\right), \label{N}\ee where $f_{ij}$ denotes the
probability entering the $i$th branch at the $j$th fork point,
thus $f_{ij}\lesssim 1$ and $\sum_i f_{ij}=1$ are required, and
${\cal N}_{ij}$ is the efolding number of inflation in which the
effective inflaton slow rolls along the $i$th branch beginning at
the $j$th fork point. Here, different paths means that for each
path, there is at least one branch, which is different from any
other path. In principle, there can be lots of paths be able to
arrive at the same minimum. Thus there is a sum for all paths,
whose end point is at certain minimum. $f_{ij}$ is generally
determined by the characters of corresponding fork point, e.g. the
number of branches and the tilt of effective potential along
different branches, and the evolutive history before it. Though
$f_{ij}\lesssim 1$, generally $e^{{\cal N}_{ij}}\gg 1$, which can
render $f_{ij}e^{{\cal N}_{ij}}\gtrsim 1$ in corresponding branch
of given path. Thus $N$ can be generally quite large.

The time that the effective inflaton spends in each branch should
be different, since the effective potential of each branch is
generally different, and it can be estimated as follows. The
efolding number for each branch is ${\cal N} =\int H_{inf} dt$,
which is given by \be {\cal N}\simeq \int {d\varphi\over M_P
\sqrt{\epsilon}}, \ee where $H_{inf}$ is the scale of inflation in
corresponding branch, which is assumed to be nearly unchanged, and
$\epsilon $ is the usual slow roll parameter. Here, $\epsilon$ is
regarded as constant for simplicity, thus the time spended in
corresponding branch is \be \Delta t\simeq {1\over H_{inf}}{\Delta
\varphi\over M_P \sqrt{\epsilon}}. \ee Thus in principle, the
smaller $\epsilon$ is, the larger this time interval is. We
require that the downhill course is not interrupted by the eternal
inflation, which corresponds to place a lower limit for $\epsilon$
in each branch, i.e. $\sqrt{\epsilon}\gtrsim {H_{inf}\over M_P}$.
Thus we have \ba \Delta t &\lesssim & {1\over
H_{inf}}\left({\Delta \varphi\over H_{inf}}\right) \simeq
\left({M_P^2\over M_{inf}^4}\right)\Delta \varphi \nonumber\\
&\simeq & \left({M_P\over M_{inf}}\right)^4 {\Delta \varphi\over
M_P}t_P, \label{dt}\ea where $\rho_{inf}=M_{inf}^4$ is the energy
density driving inflation, and $t_P={1\over M_p}\simeq {1\over
10^{43}}$s. When the inflaton potential of each branch has a
stringy or supergravity origin, the excursion of effective
inflaton in each branch can not exceed $M_p$, i.e. $\Delta \varphi
\lesssim M_p$, we have $\Delta t\lesssim ({M_P\over M_{inf}})^4
t_P$. Hence, as is an evident result, we can see that the lower
the energy scale of effective potential of a given branch is, the
longer the time that the effective inflaton experiences the
corresponding branch will be. For example, if $M_{inf}\sim 10^{15}$GeV,
$\Delta t\lesssim 10^{-31}$s can be obtained, of course if $\Delta
\varphi>M_P$ this time will be longer, while if $M_{inf}\sim
10^{3}$GeV, we have $\Delta t\lesssim 10^{21}$s, which is actually
larger than the age $10^{18}$s of our universe. The time that the
effective inflaton experiences a given path should be the sum of
the time that it experiences each branch of the corresponding
path. Thus if there are some universes which can enter into some
branches of low scale inflation, in which $M_{inf}\lesssim
10^{3}$GeV, then at present they could be still inflating. This
indicates that in this multiverse scenario, due to the diversity
of the paths in the field space, it is possible that not all
universe have thermalised, even if some universes have thermalised
for a long time.



\section{Observations of a given observer at late time}

We will show possible observations of a given observer at late
time in certain thermalised universes in this multiverse scenario.


\subsection{On primordial perturbation}

The amplitude of perturbation produced around the fork is
determined by the character of potential around this point. In
general, in a given path that appearance of new branches implies
${\partial^2 V\over
\partial \varphi_i^2}<0$ at corresponding fork point (while it is
positive before this fork point) where $\varphi_i$ denotes the
effective field along the $i$th branch.
Thus we consider the potential as \footnote{In principle, for each
branch there may be several or lots of fields participating the
slow rolling, which leads to assisted inflation \cite{LMS}, see
also \cite{MPP},\cite{PCZZ},\cite{MD}, and also Nflation
\cite{DKMW}. Recently, there have been large number of studies on
Nflation
\cite{AL},\cite{EM},\cite{Liddle},\cite{Piao0606},\cite{W},\cite{CX},
and also \cite{BBD} for staggered inflation. } \be
V(\varphi)=V_*\left(1-\beta{\varphi^2\over M_P^2}\right)
\label{p2}\ee for a detailed analysis, where $\beta$ is a positive
and dimensionless parameter. This potential is the same as that of
`hilltop inflation' \cite{BL1},\cite{KLL}, in which inflation
takes place near a maximum of the potential, thus the
perturbations generated during inflation can be suitable for
observable universe. Here, we mainly concern the perturbation
modes produced around the fork point, which may have distinct
features and can enter into the horizon of a given observer at
late time.

The parent universe is separated into different regions, or baby
universes, at the fork point `A', some of which will be dominated
by the effective inflaton along the $\varphi$ direction. The
effective mass around the fork point, i.e. $\varphi\simeq 0$,
along $\varphi$ direction is $|m_{\varphi}^2|={2\beta V_*\over
M_P^2}$. The slow roll parameter $\eta$, which together with
$\epsilon$ is generally used to depict the slow rolling required
by inflation models, is given by $|\eta|= {|m_{\varphi}^2|\over
3H^2_{inf}}\sim \beta$. In general, the amplitude of perturbation
is $\sim {1\over V^{\prime}_{\varphi}}$. Thus the amplitude seems
to be divergent at the fork point `A', since
$V^{\prime}_{\varphi}\simeq 0$. However, as has been mentioned,
the field will be generally brought out of this extremum point by its
fluctuation $\delta \varphi={H_{inf}\over 2\pi}$. Thus during the
inflation driven by $\varphi$, the amplitude of perturbation mode
initially produced is given by \be {\cal P}^{1/2}_\zeta\sim
{1\over M_P^3}{V^{3/2}\over V^{\prime}_{\varphi}}\sim
{\sqrt{V_*}\over \beta M_P \varphi_{ini}}\sim {1\over |\eta|},
\label{p2p}\ee where $\varphi_{ini}\simeq \delta
\varphi={H_{inf}\over 2\pi}$ has been used. Thus if $|\eta|> 1$
around the fork point, which may be obtained by adjusting the
parameter $\beta$ of potential (\ref{p2}), ${\cal
P}_{\zeta}^{1/2}< 1$ can be obtained  \footnote{ The result of
(\ref{p2p}) is obtained in the slow roll approximation. However,
if $H_{inf}$ is regarded as constant, the amplitude of
perturbation can be exactly generated independent of $|\eta|\ll
1$. In this case, actually the evolution of $\varphi$ can be
exactly solved, e.g.\cite{GLW}.
The amplitude of curvature perturbations around the fork point is
given by \be {\cal P}_{\zeta}^{1/2}\equiv\delta\,{\cal N}\sim
{\delta\varphi\over s\varphi}\sim {1\over s}, \ee where
$s=1.5(\sqrt{{4|\eta|\over 3}+1}-1)$, and $\varphi\simeq \delta
\varphi={H_{inf}\over 2\pi}$ has been used, and $\cal N$ is the
efolding number. 
Therefore, when $|\eta|\ll 1$, ${\cal
P}_{\zeta}^{1/2}\sim 1/|\eta|$ is consistent with Eq.(\ref{p2p});
while for $|\eta|\gtrsim 1$, ${\cal P}_{\zeta}^{1/2}\sim
1/\sqrt{|\eta|}$. }.

In general, for the potentials of new branches around the fork
point, there can be other alternatives, since only is ${\partial^2
V\over
\partial \varphi_i^2}<0$ required at corresponding fork point. We, for example, can
also consider the potential as \be
V(\varphi)=V_*\left(1-\lambda{\varphi^3\over M_P^3}\right),
\label{p3}\ee where $\lambda$ is a positive and dimensionless
parameter. The effective mass around $\varphi\simeq 0$ is
$|m_{\varphi}^2|={6\lambda \varphi V_*\over M_P^3}$, which seems
smaller than that of the potential (\ref{p2}), since $\varphi$ is
smaller. The amplitude of mode initially produced during the
inflation along this branch is given by $ {\cal
P}^{1/2}_{\zeta}\sim {1\over M_P^3}{V^{3/2}\over
V^{\prime}_{\varphi}}|_{ini}$, which is \be {\cal
P}^{1/2}_{\zeta}\sim {\sqrt{V_*}\over \lambda\varphi^2}\sim
{\sqrt{V_*}\over |\eta|M_P \varphi_{ini}}\simeq {1\over |\eta|},
\label{p3p}\ee where $|\eta|={|m_{\varphi}^2|\over 3H^2_{inf}}\sim
{\lambda \varphi\over M_P}$ and $\varphi_{ini}\simeq \delta
\varphi={H_{inf}\over 2\pi}$ have been used. Thus the case is
similar to that for the potential (\ref{p2}). However, if we
require $|\eta|\gg 1$, the potential (\ref{p3}) seems to need a
larger fine tunning, since $|\eta|\sim {\lambda \varphi/M_p}\gg 1$
means initially $\lambda \gg {M_p\over \varphi_{ini}}\sim
{M_p\over H_{inf}} $, which is certainly unnatural. However, in
principle we can naturally obtain $|\eta|> 1$ by considering other
potential, for example, the potential likes
$V_*(1+{\xi}\ln{\varphi\over \varphi_*})$, where $\xi$ is a
positive and dimensionless constant and $\varphi_*$ is certain
value at which the fork point begins. The field $\varphi$ will
roll down along the direction of $\varphi<\varphi_*$ for this
potential. The effective mass at about $\varphi\simeq \varphi_*$
is $|m_{\varphi}^2|={{\xi} V_*\over \varphi^2}$. Thus
$|\eta|={|m_{\varphi}^2|\over 3H^2_{inf}}\sim {{\xi} M_P^2\over
\varphi^2_{ini}}$. In general, $\varphi_{ini}\simeq
\varphi_*+\delta \varphi=\varphi_*$ can be smaller than $M_P$.
Thus if $\xi$ is not too small, initially $|\eta|> 1$ can be
obtained. The amplitude of mode initially produced during the
inflation along this branch is also the same as (\ref{p2p}) and
(\ref{p3p}), which is given by \be {\cal P}^{1/2}_{\zeta}\sim
{\sqrt{V_*}\varphi_{ini}\over {\xi}M_P^3}\sim {\sqrt{V_*}\over
|\eta|M_P \varphi_{ini}}\simeq {1\over |\eta|}. \ee Thus in
principle, along the directions of new branches around the fork
point, it seems that both $|\eta|\lesssim 1$ and $|\eta|>1$ are
possible, and which one is actually obtained is dependent on the
details of the effective potential around the corresponding fork
point.

We can see that for $|\eta|\lesssim 1$, the amplitude of
perturbation is generally larger than 1. This means that if the
corresponding perturbation mode enters into the horizon at late
time, it will inevitably cause gravitational collapse of
observable universe, which will rapidly gulp the observer inside a
black hole \footnote{In hybrid inflation, this will lead to the
formation of massive primordial black holes in the primordial
universe \cite{GLW}, and similar case would also happen in locked
inflation, e.g.\cite{EKS}. The problems of locked inflation has
been discussed detailed in \cite{CR}, which, however, can be
avoided in `old' locked inflation \cite{LPS}. }. Thus in this
case, it seems the observer can hardly probe the information
around the fork point \footnote{This case is similar to that
around the boundary of slow roll eternal inflation, at which
${\cal P}_\zeta\sim 1$, which in certain sense means that the
information of slow roll eternal inflation seems not accessible to
a given observer at late time, e.g.\cite{BFY}, and also
\cite{APQ},\cite{AEL} for the case with large number of fields.}.
While for $|\eta|\gg 1$, the case is different, since the
amplitude of corresponding perturbation mode is smaller than 1,
and the observer might be able to learn more about the fork point.

When $|\eta|\gg 1$, the potential along the $\varphi$ direction
around the fork point should be quite abrupt. Thus in this case
the $\varphi$ field will rapidly rolling down along its potential.
However, we have required that in each branch there is a region
suitable for a period of slow roll inflation. Thus for such case,
it seems possible that after a period of fast rolling, the
effective field will enter into a period of slow rolling, which
drives the inflation of corresponding universe. However, even if
there is such a region suitable for slow rolling in corresponding
branch, it is also uncertain whether inflation will occur in this
branch.

We can simply discuss it as follows. We assume that the potential
along $\varphi$ direction around $\varphi\simeq 0$ is step-like,
the hight of step is denoted by $\Delta V$ and the following flat
region suitable for slow roll inflation is approximately constant
denoted by $V_*$. When $\Delta V\gg V_*$, after the field switches
to this branch, the corresponding new universe will rapidly become
dominated by the kinetic energy of $\varphi$ field, since
approximately ${\dot \varphi}^2_{flat}\simeq \Delta V\gg V_*$,
where the subscript `$flat$' denotes the quantity just entering
into the flat region. During the period of kinetic domination,
${\dot\varphi}\sim 1/a^3\sim 1/t$ can be obtained. Thus during
this period the moving distance of $\varphi$ is given by \ba
\Delta \varphi &\simeq & {{\dot \varphi}_{flat} \over H_{inf}}\int
{dt\over t} \simeq {{\dot\varphi}_{flat}\over
H_{inf}}\ln{\sqrt{\Delta V\over V_*}} \nonumber\\ &\simeq &
\left(\sqrt{\Delta V\over V_*}\ln{\sqrt{\Delta V\over
V_*}}\right)M_P, \ea where $t_{flat}={1\over H_{inf}}$ and
$H_{inf}^2\simeq {V_*\over M_P^2}$ have been used. Thus we can see
that if $\Delta V\gg V_*$, generally we have $\Delta
\varphi\gtrsim M_P$. This means that only after $\varphi$ fast
rolls to certain value larger than $M_P$ might the slow roll
inflation occur in this branch. However, if $\Delta\varphi\lesssim
M_P$, as has been mentioned that if the inflation potential in
corresponding branch has a stringy or supergravity origin, the
inflation will certainly not occur in this corresponding branch,
even if there is a flat region suitable for slow rolling in this
branch \footnote{However, if there is an instant production of
particles, e.g.\cite{FKL},\cite{FKL1}, the case will be altered.
The backreaction of produced particles would induce a stage of
trapped inflation \cite{KLL1},\cite{GHSS} or just slow
the effective field down by extracting energy from it.}. However, it is
possible that the slow roll inflation might occur at following
branch after the effective inflaton passes next fork point.
Whereas it is visible that if $\Delta \varphi< M_P$, $\Delta
V\lesssim V_*$ must be required. $\Delta V\lesssim V_*$ means
${\dot \varphi}^2_{flat}\simeq \Delta V\lesssim V_*$, thus after
the field switches to the corresponding branch, the corresponding
new universe will be still inflating. In this case, we have ${\dot
\varphi}\sim e^{-3H_{inf}t}$. Thus $\Delta \varphi$ is given by
\ba \Delta \varphi &\simeq & {\dot \varphi}_{flat}\int {dt\over
e^{3H_{inf}t}}
\simeq {{\dot\varphi}_{flat}\over H_{inf}}{1\over e^{3H_{inf}t_{flat}}}\nonumber\\
& <& 0.1\left(\sqrt{\Delta V\over V_*}\right)M_P, \ea where
$t_{flat}={1\over H_{inf}}$ has been used, and the prefactor is
$1/(3e^3)$, which is smaller than 0.1. Hence we can see that if
$\Delta V\lesssim V_*$, then generally $\Delta \varphi \ll M_P$.
This result is expectant. Thus for $|\eta|\gtrsim 1$, though there
is a period of fast rolling of $\varphi$ after the fork point, if
the effective potential of corresponding branch is flat enough,
the inflation will also occur. In this case, if $\Delta V\gg V_*$,
the slow roll inflation may occur at following branch after next
fork point, while if $\Delta V\lesssim V_*$, it can occur at
current branch.


Therefore, for the observer be able to live at enough late time,
if $|\eta|\ll 1$ around the fork point, he will be rapidly gulfed
by a black hole with the scale of his current horizon, since the
amplitude ${\cal P}_{\zeta}\gtrsim 1$. While $|\eta|\gg 1$, he
will see the lower cosmic microwave background anisotropies on
large angular scales 
\footnote{This, for current observations, must occur at
corresponding time such that the corresponding scale is near the
low multipole of the CMB, or we either would not observe this
effect or it has been ruled out.}, since there is generally a
period of fast rolling preceding the slow roll inflation
\footnote{It is also possible that there will be an effective
potential with lots of small steps which would not lead to a fast
roll and thus a lower amplitude of power spectrum, but instead
lead to a modulated power spectrum and bispectrum, e.g.\cite{ACE}
for effects on the power spectrum and \cite{CEL},\cite{CEL2} for
bispectrum, which might be a better probe than a period of fast
rolling for this scenario. }, e.g.\cite{CPKL}, and the bispectrum
of fast rolling has been discussed in \cite{CEL2}. When the case
is the latter, with the continuously growing of past light cone of
the observer, the domain walls will eventually enter into his
horizon \footnote{In general, the period of inflation before
thermalization inflates the domain walls away. However, the time
that slow roll inflation lasts is finite, thus it is always
possible that at enough late time, i.e.the efolding number
required by late time evolution is equal to or larger than that
given by slow roll inflation. In this sense, one could see the
domain walls, even if they are unseen at present epoch. }.

\subsection{On domain wall}

When adjacent ${1\over H_{inf}}$ regions enter into different
branches, the domain wall will inevitably appear between them
\footnote{The domain walls may actually contribute and affect the
curvature perturbation on large angular scales, which in some
sense is similar to that of the bubble wall discussed in
\cite{G},\cite{ST},\cite{GMST},\cite{GGM},\cite{LST} for open
inflation \cite{BGT},\cite{YST}. The complete investigation on
this issue is already beyond our scope. However, it should be
pointed out that the behaviors of perturbation spectrum discussed
here are not altered qualitatively. }. However, this actually
occurs only when the adjacent regions or universes have
thermalized, or the wall will move towards the regions with larger
potential energy density, and its moving velocity is the same as
that of past light cone of the observer, thus it is impossible
that the walls enter into the horizon of a given observer. Here,
however, we will consider the case that all adjacent regions or
universes have thermalized.

In general, for domain wall, the potential is generally regarded
as $\lambda(\varphi^2-\varphi_*^2 )^2$, which actually corresponds
to the expansion of the potential (\ref{p2}) around the fork
point, where $\varphi_*=\sqrt{2\over \beta}M_P$ and
$\lambda={4V_*\beta^2\over M_P^4}$. In some region of space the
value of field is $\varphi_*$, while in adjacent region it is
minus $\varphi_*$, so there must be a region where $\varphi=0$,
which corresponds to the domain wall. The surface energy density
of domain wall is generally given by ${\cal E}\sim
{\varphi_*^2\over l}+\lambda \varphi_*^4l$, e.g.\cite{vilenkin85},
which is the sum of the contributions of the gradient term and
potential energy term, and $l$ is the thickness of the wall. The
balance between both terms gives $l\sim {1\over
\sqrt{\lambda}\varphi_*}$ and ${\cal E}\sim
\sqrt{\lambda}\varphi_*^3$.

When the thickness of the wall $l>{1\over H_{inf}}$, the interior
of wall will satisfy the conditions for inflation, thus the
topological defect inflation will inevitably occur
\cite{Linde94b},\cite{vilenkin94}. For the potential (\ref{p2}),
$l>{1\over H_{inf}}$ implies \ba &{1\over \sqrt{\lambda}\varphi_*}
& \sim \sqrt{M_P^4\over V_*\beta^2}{\sqrt{\beta}\over M_P}>
{1\over H_{inf}}\sim {M_P\over \sqrt{V_*}}\nonumber\\
&\Longrightarrow &\,\, \beta<1, \ea which means $|\eta|<1$. The
topological defect inflation is actually eternal
\cite{Linde94b},\cite{vilenkin94}, i.e. the inflation of wall will
continue eternally, even if the $\varphi$ field in some regions of
defect core $\varphi\simeq 0$ has rolled down along the potential
and entered into the corresponding branches. This is consistent
with the result of (\ref{p2p}), where for $|\eta|<1$ around the
fork point the amplitude of perturbation mode produced is ${\cal
P}_{\zeta}>1$. This is because the corresponding mode is actually
produced at the boundary of defect eternal inflation, thus its
amplitude ${\cal P}_{\zeta}>1$ is expectant. Thus in this case
even if the past light cone of the observer is enough large, he
would have been engulfed inside a black hole, before he has the
opportunity to see the wall.

However, if $|\eta|\gg 1$, the domain wall will not inflate. Thus
it is possible that at late time the domain wall will enter into
the horizon of a given observer. The domain walls are produced
during the evolution of field around the fork point. The length
scale of domain wall is initially about ${1\over H_{inf}}$, and
then is brought to $R\simeq {1\over H_{inf}}e^{\cal N}$ by
inflation in corresponding branch. Hereafter, it is stretched by
the expansion of standard cosmology. Thus given that the domain
wall enters into the horizon at present $R\sim {1\over H_0}$, its
mass can be estimated as, e.g.\cite{KT},\cite{Mukhanov}, \ba
M_{\rm wall}&\sim & {{\cal E}\over H_0^2}\sim
10^{65}\sqrt{\lambda}\left({\varphi_*\over 100{\rm
Gev}}\right)^3{\rm g}\nonumber\\ &\sim &
10^{10}\sqrt{\lambda}\left({\varphi_*\over 100{\rm Gev}}\right)^3
M_{H_0}, \ea where $M_{H_0}\sim {\rho_0\over H_0^3}$ is the mass
of observable universe at present, and $\rho_0$ is the energy
density of observable universe. Thus unless $\lambda$ and
$\varphi_*$ are unacceptablely small, $M_{\rm wall}$ is generally
larger than $M_{H_0}$. Thus when the domain wall enters into the
horizon of a given observer at late time, it will lead to a large
fluctuation of energy density in corresponding regions. Here,
relevant phenomena might be significant, which will be left for
future works.



\section{Conclusion}

In a given path with multiple branches, each of which may drive a
period of slow roll inflation, it can be expected
that there are some fork points, at which one branch is bifurcated
into different branches, or many branches converge into one or
several branches. In this paper, it is showed that if there is a
web of such branches in a given field space, a multiverse
separated by domain wall network will actually come into being.

Rolling down hill, the effective field will be expected to
enter into certain stable or metastable minimum, which corresponds
to the terminal of a given path. In principle, there should be
large number of such minima, some of which might be suitable for
the world where we live. We estimated the number of universes
entering into certain terminal minimum, which is generally quite
large. For a given observer be able to live permanently, he might
be able to explore the modes produced around the fork point. In
this case, if $|\eta|\ll 1$ along the direction of effective
inflaton around the fork point, he will be gulfed by a
black hole with the scale of his current horizon rapidly, while if
$|\eta|\gg 1$, he will see the lower cosmic microwave background
anisotropies on large angular scales, since there is generally a
period of fast rolling preceding the slow roll inflation. When the
case is the latter, at enough late time, the domain walls will
possibly enter into the horizon of this given observer, which
would bring a distinct observation.

In general, the multiverse scenario can be obtained in eternal
inflation, which is based on the random walk of inflaton field
induced by itself large quantum fluctuation.
Here, however, the multiverse scenario is implemented by the
classical rolling of inflaton along a web of branches of its
effective potential which happens at a relatively
lower energy scale. In reality to the multiverse scenario, due to
the complexity of potential of field space, both effects may be
possible to contribute the number of universes, and
the possibility that a region of space evolving along one branch
classically tunnels to another close-by branch is also not excluded.
Thus the
exploration of relevant issues will be interesting. In addition,
it is also significant to investigate a possible implement of such
a scenario in string theory, which will be studied in the future.

\textbf{Acknowledgments} We thank T. Battefeld, Y.F. Cai, Y. Wang
for discussions. This work is supported in part by NSFC under
Grant No: 10775180, in part by the Scientific Research Fund of
GUCAS, in part by CAS under Grant No: KJCX3-SYW-N2, in part by the
Ministry of Science and Technology of China under Grant
No:2010CB832800.

\end{document}